\documentstyle[preprint,aps]{revtex}

\tighten
\input epsf

\begin{document}
\draft

\title{\Large \bf Deterministic Equations of Motion 
and Dynamic Critical Phenomena}

\author{\bf B. Zheng$^{*\dagger}$, M. Schulz$^*$ and 
S. Trimper$^*$}

\address{$^*$ Universit\"at -- Halle, 06099 Halle, 
Germany}

\address{$^\dagger$ Universit\"at -- GH Siegen, 57068 Siegen, 
Germany}

\maketitle

\begin{abstract}
Taking the two-dimensional $\phi^4$ theory as an example,
we numerically solve the deterministic equations
of motion with random initial states. 
Short-time behavior of the solutions is systematically
investigated.
Assuming that the solutions 
 generate a microcanonical
ensemble of the system, we demonstrate that
 the second order phase transition
point can be determined
already from the short-time dynamic behavior.
Initial increase of the magnetization and critical slowing down
are observed.
The dynamic critical exponent $z$, the new exponent $\theta$ and
the static exponents $\beta$ and $\nu$ are estimated.
Interestingly, the deterministic dynamics with random
initial states is in a same dynamic universality class
of Monte Carlo dynamics.
\end{abstract}


It is believed that statistical mechanics 
is originated from the
fundamental equations of motion
for many body systems or field theories, even though up to now 
there exists
not a general proof. For classical systems, equations of
motion are  {\it deterministic}. Statistical
ensembles are expected to be effective description of the systems.
For quantum systems the situation is similar.
In any case, practically it remains open
whether solutions of the fundamental equations of motion really yield
the same results as statistical mechanics, e.g., see Refs.
\cite {fer65,for92,esc94,ant95,els97}.

With recent development of computers, 
to solve numerically equations of motion
gradually becomes possible. 
This has attracted much attention of scientists in different areas.
For example, recently the $O(N)$ vector model
 and $XY$ model have been numerically solved
 \cite {cai98,cai98a,leo98}.
Noting that energy is conserved during time evolution,
 solutions of equations of motion
actually generate a {\it microcanonical} ensemble.
Making the time average of the observables,
introducing standard techniques developed in statistical mechanics
for the canonical ensemble,
phase transition points and critical exponents
have been estimated.
The results are in agreement with those obtained
from a canonical ensemble in statistical mechanics.

In principle, the deterministic 
equations of motion should describe not only
equilibrium properties but also dynamic properties of the systems.
In statistical mechanics, dynamics is approximately given
by some effective {\it stochastic} equations of motion, typically at
mesoscopic level, e.g. Langevin equations or Monte Carlo
algorithms. For stochastic dynamics, critical slowing down
and dynamic scaling around the critical point are characteristic
properties. 
It has long been challenging whether stochastic dynamics
describes correctly the real physical world.
It is important and interesting to study
dynamic properties of the fundamental deterministic 
equations of motion and whether deterministic dynamics 
and stochastic dynamics are in a same universality class.

On the other hand, in recent years much progress has been achieved
in stochastic dynamics. For a long time it is known that
around the critical point there exists universal scaling behavior 
in the long-time regime of the dynamic evolution.
This is more or less due to the divergent correlation time
 induced by the divergent spatial correlation
length. Short-time behavior is believed to depend on
microscopic detail. However, in recent years it is discovered that
universal scaling behavior emerges already in the 
{\it macroscopic} short-time regime, after a time scale
$t_{mic}$ which is sufficiently large in the microscopic sense
\cite {jan89,hus89,sta92,oer93,li94,maj96a,luo98}
(for a review, see Ref.~\cite {zhe98}).
Important is that one should take care of the macroscopic 
initial conditions carefully.
For example, dynamic magnetic systems with an initial state
of very high temperature and small magnetization,
the short-time dynamic scaling
for the $k$th moment of the magnetization is written as \cite {jan89}
\begin{equation}
M^{(k)}(t,\tau,L,m_{0})=b^{-k\beta/\nu}
M^{(k)}(b^{-z}t,b^{1/\nu}\tau,b^{-1} L,
b^{x_{0}}m_{0}).
\label{e10}
\end{equation}
Here $\tau=(T-T_c)/T_c$ is the reduced temperature, 
 $\beta$, $\nu$ are the well known static critical exponents
 and $z$ is the dynamic exponent,
while the {\it new independent} exponent $x_0$ 
is the scaling dimension
of the initial magnetization $m_0$.
The parameter $b$ is an arbitrary rescaling factor
and $L$ is the lattice size.
For the dynamics of model A, a prominent property of
the short-time dynamic scaling is that the magnetization
undergoes a {\it critical initial increase} at early time
\cite {jan89,li94,gra95,sch95}
\begin{equation}
M(t) \sim m_0 \, t^\theta
\label{e20}
\end{equation}
where the exponent $\theta$ is related to $x_0$ by
$\theta =(x_0 - \beta /\nu)/z$.

More interesting and important is that the static exponents
$\beta$, $\nu$ and the dynamic exponent $z$ 
in the short-time scaling form (\ref {e10}) take the same values as they are
defined in equilibrium or the long-time regime.
This fact makes it possible to determine
all these exponents already in the short-time regime
of the dynamic evolution \cite {zhe98}.
 Since now the measurements are carried out
at early time, the method is free of critical slowing down.
The short-time dynamic scaling is not only conceptually interesting 
but also practically important.
To clarify whether the short-time dynamic scaling
is a fundamental phenomenon of the real physical world
becomes urgent.

The purpose of this letter is to study whether
there exists short-time universal scaling behavior
in the dynamics described by the deterministic equations of motion
and meanwhile determine the critical point and
all the static and dynamic exponents.

The model we choose is the two-dimensional $\phi^4$ theory.
The reason is that the statics
of this model is known in a same universality class
of the two-dimensional Ising model and 
there exist already some numerical results
 \cite {cai98}.
The Hamiltonian of the model on a square lattice is
\begin{equation}
H=\sum_i \left [ \frac{1}{2} \pi_i^2 
  + \frac{1}{2} \sum_\mu (\phi_{i+\mu}-\phi_i)^2 
  - \frac{1}{2} m^2 \phi_i^2 
  + \frac{1}{4!} \lambda \phi_i^4 \right ]
\label{e30}
\end{equation}
with $\pi_i=\dot \phi_i$ and it leads to the equations of motion
\begin{equation}
 \ddot \phi_i= \sum_\mu (\phi_{i+\mu}+\phi_{i-\mu}- 2\phi_i)
  +  m^2 \phi_i
  - \frac{1}{3!} \lambda \phi_i^3\ .
\label{e40}
\end{equation}

In the dynamic evolution governed by
 equations (\ref {e40}), energy is conserved.
The solutions are assumed to generate a microcanonical ensemble.
The temperature could not be introduced 
externally as in a canonical ensemble,
but could only be defined internally as the averaged
kinetic energy. In our short-time dynamic approach,
the total energy is actually  a even more convenient controlling parameter
of the system, since it is conserved and 
can be input from the initial state. 
Therefore, from now $\tau$ in the scaling form (\ref {e10})
will be understood as a reduced energy density 
$(\epsilon-\epsilon_c)/\epsilon_c$.
Here $\epsilon_c$ is the critical energy density corresponding to 
a second order phase transition.

The order parameter of the $\phi^4$ theory is the magnetization.
The time-dependent magnetization $M\equiv M^{(1)}(t)$  and its second moment
 $M^{(2)}$ are defined as
\begin{equation}
M^{(k)}(t)=\frac {1}{L^{2k}} 
\langle \left [ \sum_i \phi_i(t) \right ]^{(k)} \rangle, \quad k=1, 2.
\label{e50}
\end{equation}
The average for observables is {\it only
over initial configurations}. 
In the short-time regime of the dynamic evolution,
spatial correlation length is small, one can easily realize
that $M^{(2)} \sim 1/L^d$. From the scaling form 
(\ref {e10}), we obtain at the critical point for $m_0=0$
\cite {hus89,sch95,zhe98}
\begin{equation}
M^{(2)}(t) \sim t^{(d-2\beta/\nu)/z}.
\label{e60}
\end{equation}

Another interesting observable is the auto-correlation function
\begin{equation}
A(t)=\frac {1}{L^2} \sum_i < \phi_i(0) \phi_i(t) >.
\label{e70}
\end{equation}
From the calculations based on Langevin equations,
at the critical point the auto-correlation $A(t)$ with $m_0=0$
presents a power law behavior \cite {jan92}
\begin{equation}
A(t) \sim t^{-d/z + \theta}.
\label{e80}
\end{equation}
The power law behavior indicates {\it critical slowing down}.
The interesting point is that the new exponent $\theta$
 describing the initial increase of the magnetization
 also appears here. 

Our strategy is that from initial increase of the magnetization,
Eq.~(\ref {e20}), we measure the exponent $\theta$. 
Taking $\theta$ as an input, from the power law decay of
the auto-correlation, Eq.~(\ref {e80}),
 we extract the dynamic exponent $z$.
 Then from power law behavior of the second moment,
Eq.~(\ref {e60}), we obtain the static exponent $\beta/\nu$.
Finally, the critical point and the remaining
exponent $\nu$ are determined as follows.
Assuming $L$ is sufficiently large and $t^{x_0/z}$ is small enough,
one can deduce from scaling form (\ref {e10})
\begin{equation}
M(t)=m_0t^{\theta}F(t^{1/\nu z} \tau).
\label{e90}
\end{equation}
When $\tau=0$, Eq.~ (\ref {e20}) is recovered.
When  $\tau \neq 0$, the power law behavior will be modified.
Therefore, searching for a energy density
which gives the best power law behavior for the magnetization,
one determines the critical point \cite {zhe98}.
The exponent $1/\nu z$ can be extracted from 
the logarithmic derivative \cite {zhe98}
\begin{equation}
\partial_\tau \ \ln \ M(t,\tau,m_{0})|_{\tau=0} \sim  t^{1/\nu z}.
\label{e100}
\end{equation}

For stochastic
dynamics in a canonical ensemble, to obtain the scaling form
(\ref {e10}) and (\ref {e80}) 
the initial state is taken to be a very high temperature
state.
Therefore, for deterministic dynamics we consider
a {\it random} initial state 
with zero or small magnetization.
For simplicity, we also set initial kinetic energy to be
zero, i.e. $\dot \phi_i(0)=0$.
Practically we proceed as follows. We first fix
the magnitude of the initial field to be a constant $c$, 
$|\phi_i(0)|=c$, then randomly give the sign to $\phi_i(0)$
with the restriction of a fixed magnetization in unit of $c$,
and finally the constant $c$ is determined by
the given energy.
Following Ref. \cite {cai98}, we take parameters $m^2=2.$
and $\lambda=0.6$.

To solve these equations of motion numerically,
 we simply discretize
$\ddot \phi_i$ by 
$(\phi_i(t+\Delta t)+\phi_i(t-\Delta t)-2\phi_i(t))/(\Delta t)^2$.
After an initial configuration
is prepared,
we update the equations of motion until
$t=500$. Then we repeat the procedure with another
initial configuration.
In our calculations, we use 
a fairly large lattice $L=256$ and samples
of initial configurations
for average range from $4\ 500$ to $10\ 000$
depending on initial magnetization $m_0$.
Errors are simply estimated by dividing total samples
into two or three groups.
Extra calculations with $L=128$ show
that the finite size effect for $L=256$
is already negligible small.

Compared with numerical solutions
in the long-time regime, the finite $\Delta t$ effect
in the short-time dynamic approach is less severe,
since our updating time is limited.
Our results are presented
with $\Delta t =0.05$. We have also performed
some simulation with $\Delta t =0.01$
and confirmed that the finite $\Delta t$ effect
has been rather small. 
In any case, however, since we have this small $\Delta t$,
the calculations are much more lengthy
than standard Monte Carlo simulations.
Therefore, the purpose of this paper is 
to explore new physics rather than 
to pursue
high accuracy of measurements.

\begin{figure}[p]\centering
\epsfysize=6.cm
\epsfclipoff
\fboxsep=0pt
\setlength{\unitlength}{0.6cm}
\begin{picture}(9,9)(0,0)
\put(-2,-0.5){{\epsffile{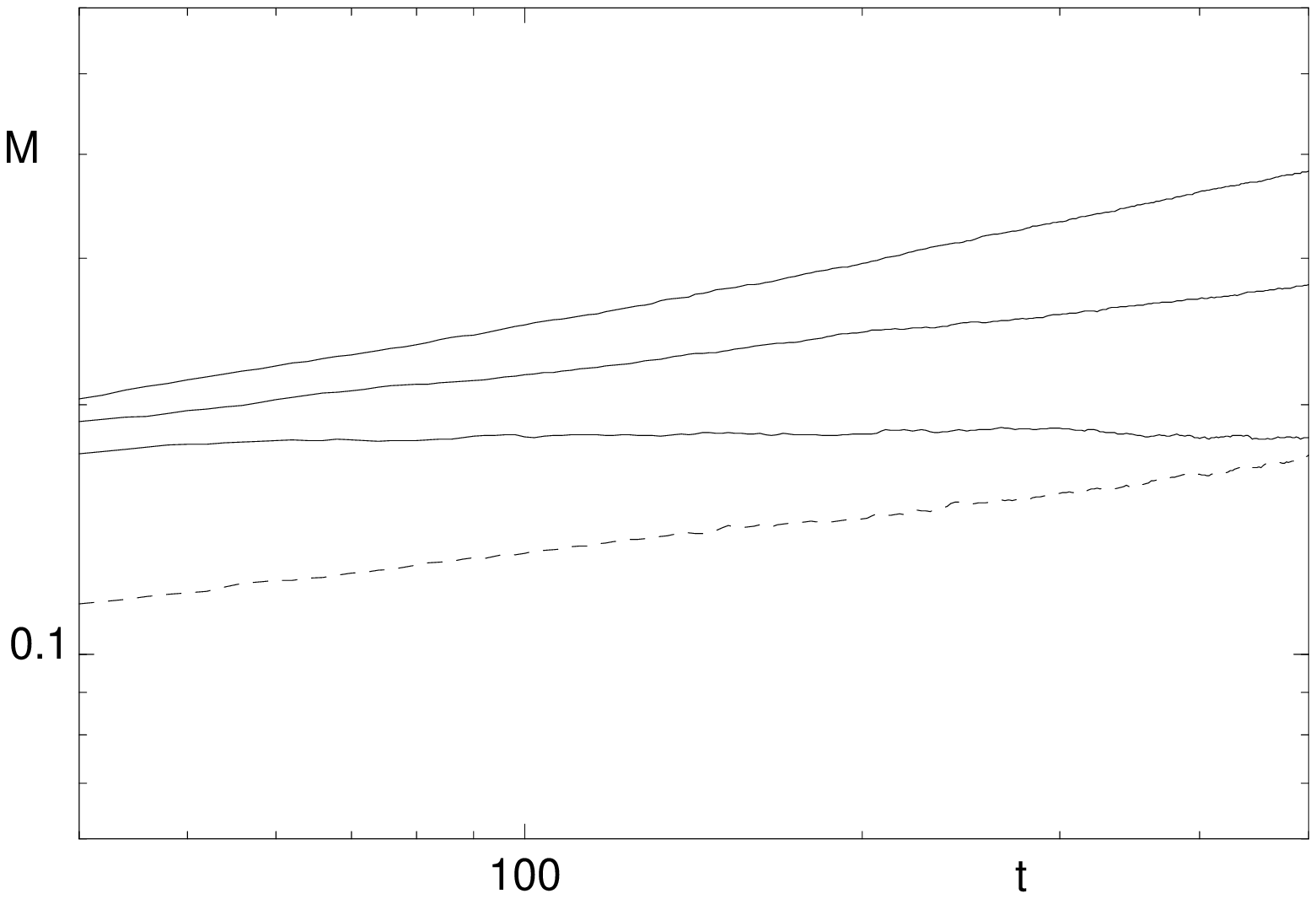}}}
\end{picture}
\caption{Time evolution of the magnetization in log-log scale. 
Solid lines are for $m_0=0.015$ with energy densities
$\epsilon=20.7$, $21.1$ and $21.5$ (from above),
while the dashed line is for $m_0=0.009$ with $\epsilon_c=21.11$.
}
\label{f1}
\end{figure}

In order to determine the critical point, 
we perform simulations with non-zero
initial magnetization $m_0$ for
a couple of energy densities in the critical regime.
In Fig.~\ref {f1}, time evolution of the magnetization
with $m_0=0.015$
for parameters  $m^2=2.$
and $\lambda=0.6$
has been plotted with solid lines for three energy densities
$\epsilon=20.7$, $21.1$ and $21.5$
in log-log scale. To show the universal behavior clearly, 
we have cut the data for $t < t_{mic} \approx 50$. 
Even looking by eyes, one could realize that the magnetization of 
$\epsilon=21.1$ has rather good power law behavior.
Actually, careful analysis of the data between $t=50$ and $500$
leads to the critical energy density
 $\epsilon_c=21.11(3)$.
 This agrees well with $\epsilon_c=21.1$ given 
 by the Binder cumulant in
 equilibrium in Ref.~\cite {cai98}.

 At $\epsilon_c$, one measures the exponent $\theta=0.146(3)$.
 Accurately speaking, however, the exponent $\theta$ is defined
 in the limit $m_0 \rightarrow 0$.
 In general, for finite $m_0$ the exponent $\theta$ may show
 some $m_0$-dependence \cite {zhe98}. Therefore, we have performed
 another simulation with $m_0=0.009$ at $\epsilon_c$.
 The magnetization is also displayed in Fig.~\ref {f1}
 with a dashed line. The corresponding exponent is $\theta=0.158(2)$.
 If we linearly extrapolate the results
 to $m_0=0$, we obtain the final value $\theta=0.176(7)$.
 
\begin{figure}[p]\centering
\epsfysize=6.cm
\epsfclipoff
\fboxsep=0pt
\setlength{\unitlength}{0.6cm}
\begin{picture}(9,9)(0,0)
\put(-2,-0.5){{\epsffile{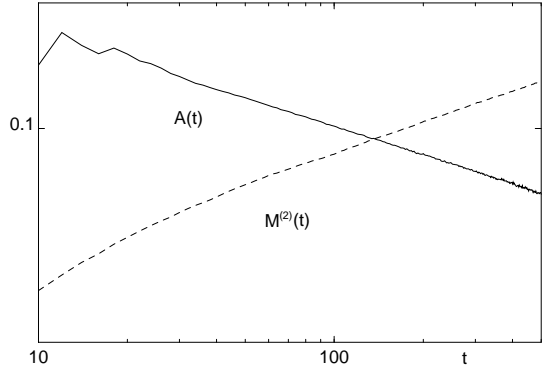}}}
\end{picture}
\caption{The auto-correlation $A(t)$ (solid line) and second moment
$M^{(2)}(t)$ (dashed line) with $m_0=0$ at 
the critical point in log-log scale.}
\label{f2}
\end{figure}

 With the critical energy in hand, we set $m_0=0$ and proceed to
 measure the auto-correlation $A(t)$ and second moment $M^{(2)}(t)$ .
 In Fig.~\ref {f2}, $A(t)$ and
 $M^{(2)}(t)$ are displayed
 in log-log scale with a solid and dashed line
 respectively. In order to show some typical microscopic
 behavior within $t_{mic}$, here we present the data from 
 a relatively early time $t=10$. 
 From the figure we see clearly that at very early time,
 the curves do not have power law behavior, typically oscillating a bit.
 However, after $t > t_{mic} \approx 50$, both curves
 nicely stabilize to power law behavior.
 From the data for $t > 50$, we measure the exponents
 $d/z-\theta=0.755(5)$ and
 $(d-2\beta/\nu)/z=0.819(12)$.

\begin{figure}[p]\centering
\epsfysize=6.cm
\epsfclipoff
\fboxsep=0pt
\setlength{\unitlength}{0.6cm}
\begin{picture}(9,9)(0,0)
\put(8.,4.){\makebox(0,0){\footnotesize $\partial_\tau \ \ln \ M$}}
\put(-2,-0.5){{\epsffile{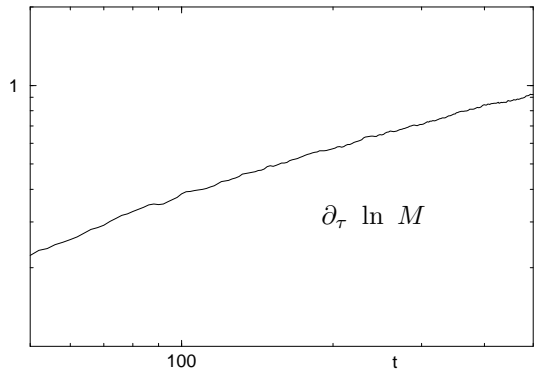}}}
\end{picture}
\caption{Logarithmic derivative of the magnetization at
the critical point in log-log scale. 
}
\label{f3}
\end{figure}

Finally, to complete the estimate of the exponents,
we calculate approximately the derivative of the magnetization
with respect to the energy density with the data
for Fig.~\ref {f1}.
In Fig.~\ref {f3}, the logarithmic derivative
of the magnetization is plotted in log-log scale
at the critical point. As usual, the fluctuation here is larger
\cite {zhe98}.
The power law behavior is also not as clean as that of
the previous observables. 
The slope of the curve tends to be smaller as the time $t$ evolves.
To improve this situation, probably we need to perform
simulations with energy densities closer to 
$\epsilon_c$ and with high statistics
and/or to consider corrections to scaling.
This requests very much computer times.
In any case, from the slope
between $t=200$ and $500$, we estimate
$1/\nu z=0.492(26)$.

In Table~\ref {t1}, we summarize all the values
of the critical exponents discussed above.
From these values, we can estimate the critical
exponents $z$, $2\beta/\nu$ and $\nu$.
The results are also listed in Table~\ref {t1}.
For comparison, the exact values of $2\beta/\nu$ and $\nu$
for the Ising model and available results for other
exponents measured from
a similar dynamic process
with standard Monte Carlo dynamics
 are also given in Table~\ref {t1}.
Remarkably, not only the static exponents, but also
the dynamic exponents of the $\phi^4$ theory
are very close to those of the Ising model
 with standard Monte Carlo dynamics.
 The exponent $\theta$ shows some percents
 of difference. However, taking into account that
 the measurements of $\theta$ are not so easy,
 especially the extrapolation to $m_0=0$ limit
 may induce some systematic errors,
 we would consider that 
 the exponent $\theta$ for both model are
 the same. Therefore, we conclude that 
 the $\phi^4$ theory described by the 
 deterministic equations of motion are
 in a same static as well as dynamic universality class
 of the Ising model with standard Monte Carlo dynamics.

Why is the deterministic dynamics in a same universality class
of the stochastic dynamics? The origin may be traced 
back to the random initial state and the kinetic energy
term in the Hamiltonian. The random initial configuration
of $\{\phi_i\}$
induced stochastically evolving kinetic energy.
The stochastic kinetic energy serves as a kind of noises
or heat bath
to the potential energy. 
This is similar to
the stochastic dynamics
with a canonical ensemble,  where the Hamiltonian is simply
the potential energy term here.

In conclusions, we have numerically solved the deterministic
 equations of motion for the two-dimensional
$\phi^4$ theory. Short-time universal behavior
is found. 
Based on the short-time dynamic scaling form,
the critical point and all the static and dynamic
critical exponents are determined.
The values of the static exponents 
agree with those of the Ising model.
More interestingly,
 both the dynamic exponents $z$ and $\theta$
also coincide 
with those of the kinetic Ising model
induced by the standard Monte Carlo algorithms
(heat-bath and Metropolis et al).
Especially, the dynamic exponent $z=2.15(2)$ is very far from $z=1$,
which is the expectation from the naive deterministic viewpoint.
These results show that, on the one hand,
the deterministic equations of motion indeed can describe
statistical properties of the systems both in statics and dynamics,
on the other hand, the Monte Carlo dynamics can be 
a good effective description of the real physical dynamics,
at least in some cases. 

Since the measurements are carried out in the short-time
regime of the dynamic evolution, our dynamic approach 
does not suffer from critical slowing down.
Furthermore, the errors induced by a finite $\Delta t$
in our measurements are also limited.
It is challenging to derive analytically the short-time dynamic
scaling from deterministic equations of motion.
Extension of the present work to quantum systems
would be very interesting.

{\bf Acknowledgements}:
Work supported in part by the Deutsche Forschungsgemeinschaft;
Schu 95/9-1 and SFB~418.

\begin{table}[h]\centering
\begin{tabular}{cccccccc}
       & $\theta$  & $d/z-\theta$ & $(d-2\beta/\nu)/z$ & $1/\nu z$
                 & $z$          & $2\beta/\nu$       & $\nu$     \\
$\phi^4$ & 0.176(7)& 0.755(5)   & 0.819(12)        & 0.492(26)
                 & 2.148(20)  & 0.24(3)        &  0.95(5) \\
Ising  & 0.191(1)& 0.737(1)   & 0.817(7)         &
                 & 2.155(3)   & 1/4              & 1\\
\end{tabular}
\caption{The critical exponents measured for the $\phi^4$ theory
in comparison with those of the Ising model.
The values of $2\beta/\nu$ and $\nu$ for the Ising model
are exact, while others are taken from Table 2 in
Ref.~\protect\cite {zhe98}.
}
\label{t1}
\end{table}


\begin{thebibliography}{10}

\bibitem{fer65}
{E. Fermi, J. Pasta and S. Ulam},  in {\em {Collected Papers of Enrico Fermi}},
  edited by {E. Segr\'e} (Univ. Chicago, Chicago, 1965).

\bibitem{for92}
{J. Ford}, Phys. Rep. {\bf {213}},  271  (1992).

\bibitem{esc94}
{D. Escande, H. Kantz, R. Livi and S. Ruffo}, J. Statist. Phys. {\bf {76}},
  605  (1994).

\bibitem{ant95}
{M. Antoni and S. Ruffo}, Phys. Rev. {\bf {E52}},  2361  (1995).

\bibitem{els97}
{Y. Elskens and M. Antoni}, Phys. Rev. {\bf {E55}},  6575  (1997).

\bibitem{cai98}
{L. Caiani, L. Casetti and M. Pettini}, J. Phys. {\bf {A31}},  3357  (1998).

\bibitem{cai98a}
{L. Caiani, L. Casetti, C. Clementi, G. Pettini, M. Pettini and R. Gatto},
  Phys. Rev. {\bf {E57}},  3886  (1998).

\bibitem{leo98}
{X. Leoncini and A.D. Verga}, Phys. Rev. {\bf {E57}},  6377  (1998).

\bibitem{jan89}
{H. K. Janssen, B. Schaub and B. Schmittmann}, Z. Phys. {\bf {B 73}},  539
  (1989).

\bibitem{hus89}
D.~A. Huse, Phys. Rev. {\bf {B 40}},  304  (1989).

\bibitem{sta92}
D. Stauffer, Physica {\bf {A 186}},  197  (1992).

\bibitem{oer93}
K. Oerding and H.~K. Janssen, J. Phys. {\bf {A26}},  3369,5295  (1993).

\bibitem{li94}
{Z.B. Li, U. Ritschel and B. Zheng}, J. Phys. {\bf {A27}},  L837  (1994).

\bibitem{maj96a}
{S.N. Majumdar, A.J. Bray, S. Cornell and C. Sire}, Phys. Rev. Lett. {\bf
  {77}},  3704  (1996).

\bibitem{luo98}
{H.J. Luo, L. Sch\"ulke and B. Zheng}, Phys. Rev. Lett. {\bf {81}},  180
  (1998).

\bibitem{zhe98}
B. Zheng, Int. J. Mod. Phys. {\bf B12},  1419  (1998), review article.

\bibitem{gra95}
P. Grassberger, Physica {\bf {A 214}},  547  (1995).

\bibitem{sch95}
L. {Sch\"ulke} and B. Zheng, Phys. Lett. {\bf {A 204}},  295  (1995).

\bibitem{jan92}
{H. K. Janssen},  in {\em {From Phase Transition to Chaos}}, edited by {G.
  Gy\"orgyi, I. Kondor, L. Sasv\'ari and T. T\'el, Topics in Modern Statistical
  Physics} (World Scientific, Singapore, 1992).

\end{thebibliography}
\end{document}